\documentclass[uplatex]{article}

\usepackage[dvipdfmx]{graphicx}
\usepackage{wrapfig}
\graphicspath{{../Picture1/}}

\usepackage{multirow}
\usepackage{here}

\makeatletter
\newcommand{\figcaption}[1]{\def\@captype{figure}\caption{#1}}
\newcommand{\tblcaption}[1]{\def\@captype{table}\caption{#1}}
\makeatother

\usepackage{longtable}

\usepackage{enumitem}

\usepackage{color}

\makeatletter
\def\Hline{\noalign{\ifnum0=`}\fi\hrule\@height 3.\arrayrulewidth \futurelet\reserved@a\@xhline}
\makeatother

\newcommand{\ctext}[1]{\raise0.2ex\hbox{\textcircled{\scriptsize{#1}}}}

\usepackage[top=30truemm,bottom=30truemm,left=20truemm,right=20truemm]{geometry}
\usepackage{fancyhdr}
\usepackage{amsmath}
\usepackage{amssymb}
\usepackage{mathcomp}
\usepackage{mathtools} 
\usepackage{cases}
\usepackage{cancel}
\usepackage{empheq}

\usepackage{comment} 
\usepackage{authblk}
\author[1,*]{Shoki Iwaguchi}
\affil[1]{\footnotesize Department of Physics, Nagoya University, Nagoya, Aichi, 464-8602, Japan}
\author[2]{Atsushi Nishizawa}
\affil[2]{ Research Center for the Early Universe (RESCEU),
Graduate School of Science, The University of Tokyo, Tokyo 113-0033, Japan}
\author[3]{Yanbei Chen}
\affil[3]{Theoretical Astrophysics 350-17, California Institute of Technology, Pasadena, California 91125, USA}
\author[1]{Yuki Kawasaki}
\author[5,1]{Masaaki Kitaguchi}
\affil[5]{The Kobayashi-Maskawa Institute for the Origin of Particles and the Universe, Nagoya University, Nagoya, Aichi 464-8602, Japan}
\author[1]{Taigen Morimoto}
\author[1]{Tomohiro Ishikawa}
\author[1]{Bin Wu}
\author[1]{Izumi Watanabe}
\author[1]{Ryuma Shimizu}
\author[1,5]{Hirohiko Shimizu}
\author[4]{Yuta Michimura}
\affil[4]{Department of Physics, University of Tokyo, Bunkyo, Tokyo 113-0033, Japan \normalsize}
\author[1,5]{Seiji Kawamura}

\affil[*]{Correspondence: iwaguchi\_s@u.phys.nagoya-u.ac.jp; +81-052-789-5982}

\title{Displacement-noise-free neutron interferometer for gravitational wave detection using a single Mach-Zehnder configuration}

\date{}

\begin{document}
\maketitle
\begin{abstract}
\noindent
The improvement of sensitivity to gravitational waves (GWs) at lower frequencies is still challenging on account of displacement noise. One of the solutions is the neutron displacement-noise-free interferometer (DFI). We focus on a simplification of the detector configuration by taking advantage of the ability to adjust the neutron speed depending on the configuration. The new configuration consists of two beamsplitters and two mirrors, which constitute a single Mach-Zehnder interferometer (MZI). It is simpler than the configuration with two MZIs in previous research. All displacement noise of mirrors and beamsplitters can be canceled in the frequency domain. This cancellation can be explained intuitively using a phasor diagram.
\\
\\
\noindent
{\em keywords:}
Gravitational wave ; Neutron interferometer ; Displacement-noise free interferometer ; Mach-Zehnder interferometer
\end{abstract}




\section{Introduction}

\ \ \ \ Ground-based gravitational wave (GW) detectors, such as LIGO and Virgo \cite{LIGO}\cite{Virgo}, have contributed to the development of GW observations. These detectors have already detected 90 GW signals from compact binary coalescences \cite{LIGO and Virgo}, and Einstein Telescope \cite{ET} and Cosmic Explorer \cite{CE} were proposed as next-generation ground-based GW detectors. Improvement of the sensitivity at lower frequencies is still challenging on account of displacement noise sources, such as thermal noise, seismic noise, and radiation pressure noise. One of the solutions for this problem is observation with space-based GW detectors, which are free from seismic noise, to increase the sensitivity at lower frequencies. For example, LISA \cite{LISA} and DECIGO \cite{DECIGO_1}\cite{DECIGO_2} are planned as space-based GW detectors. However, developing GW detectors for space is expensive and time consuming. Therefore, it is crucial to reduce displacement noise significantly for the ground-based detectors, because GWs at lower frequencies are important science targets. For example, detection of primordial GWs will probably enable us to determine which cosmic inflation model is correct.

One of the ideas to remove this obstacle to observations at lower frequencies is the displacement-noise-free interferometer (DFI), which was proposed in \cite{DFI_3}. The DFI is based on the idea that, in the transverse-traceless gauge of a GW, GW perturbations can be distinguished from displacement perturbations \cite{DFI_1}\cite{DFI_2}. The DFI signal is composed of an appropriate combination of several interferometer signals. This combination can remove displacement noise while maintaining the GW signals \cite{DFI_4}\cite{DFI_5}. At frequencies lower than 1Hz, however, the DFI has less sensitivity to GWs because the propagation time of light is much shorter than the period of the GWs. When interferometer signals are combined to cancel displacement noise at these low frequencies, the combination also cancels most of the GW signals. The DFI has highest sensitivity to GWs with periods comparable to the light propagation time in the DFI. For instance, an interferometer with arm lengths of $3\times10^8$ m has the highest sensitivity to the GWs at 1Hz, yet this long arm length is not practical for ground-based detectors. For this problem, various options such as Fabry-Perot cavities have been considered, but they have proven to be problematic \cite{FPcavity}. To resolve this problem, DFI with neutrons, which is called a \textit{neutron displacement-noise-free interferometer}, was proposed in \cite{DFNI_Nishizawa}. In a neutron DFI with neutrons propagating much more slowly than light, the neutron propagation time can be comparable to the period of GWs at lower frequencies. This enables us to cancel displacement noise without cancellation of the GW signals. Accordingly, a neutron DFI has high sensitivity to GWs at lower frequencies.

In this paper, we discuss simplification of the configuration of a neutron DFI. Generally, a simpler detector configuration is better from various perspectives. For example, a detector that has more components experiences more alignment difficulty. Furthermore, a detector that has a more complicated mechanical configuration suffers from more structural obstacles such as mechanical resonances. For these reasons, we focus on a simplification of the detector configuration by taking advantage of adjusting the neutron speed. We discuss the neutron DFI configuration in Section \ref{sec:2.1}, the cancellation of mirror displacement noise in Section \ref{sec:2.2}, the cancellation of beamsplitter displacement noise at Section \ref{sec:2.3}, a phasor diagram of the noise cancellation in Section \ref{sec:2.4}, the neutron trajectory in the neutron DFI in Section \ref{sec:2.5}, and finally the GW response in a neutron DFI in Section \ref{sec:3}.

\section{Neutron DFI using a single Mach-Zehnder configuration with two pairs of bidirectional neutrons at different speeds}

\subsection{Configurations of neutron DFIs}
\label{sec:2.1}

The concept of the neutron DFI is based on the laser DFI \cite{DFI_3}. The straightforward neutron DFI has the configuration shown in Figure \ref{fig:1} (a). In this configuration, a pair of counter-propagating neutrons comprises one Mach-Zehnder interferometer (MZI). In this paper, a pair of counter-propagating neutrons is called a “bidirectional neutron.” Two bidirectional neutrons, which are four counter-propagating neutrons with the same speed, comprise configuration (a), which is composed of One large MZI and one small MZI. Instead of a laser, this configuration uses neutrons for improving the sensitivity to GWs at lower frequencies \cite{DFNI_Nishizawa}. For simplifying the  neutron DFI configuration, configuration (a) is modified to configuration (b). The large and small MZIs using two bidirectional neutrons with the same speed are replaced with a single MZI using two bidirectional neutrons with slow and fast speeds. This neutron DFI configuration is possible with neutrons, but not possible with laser light because the speed of neutrons can be adjusted arbitrarily.

\begin{figure}[h]
    \centering
    \includegraphics[clip,height=6cm]{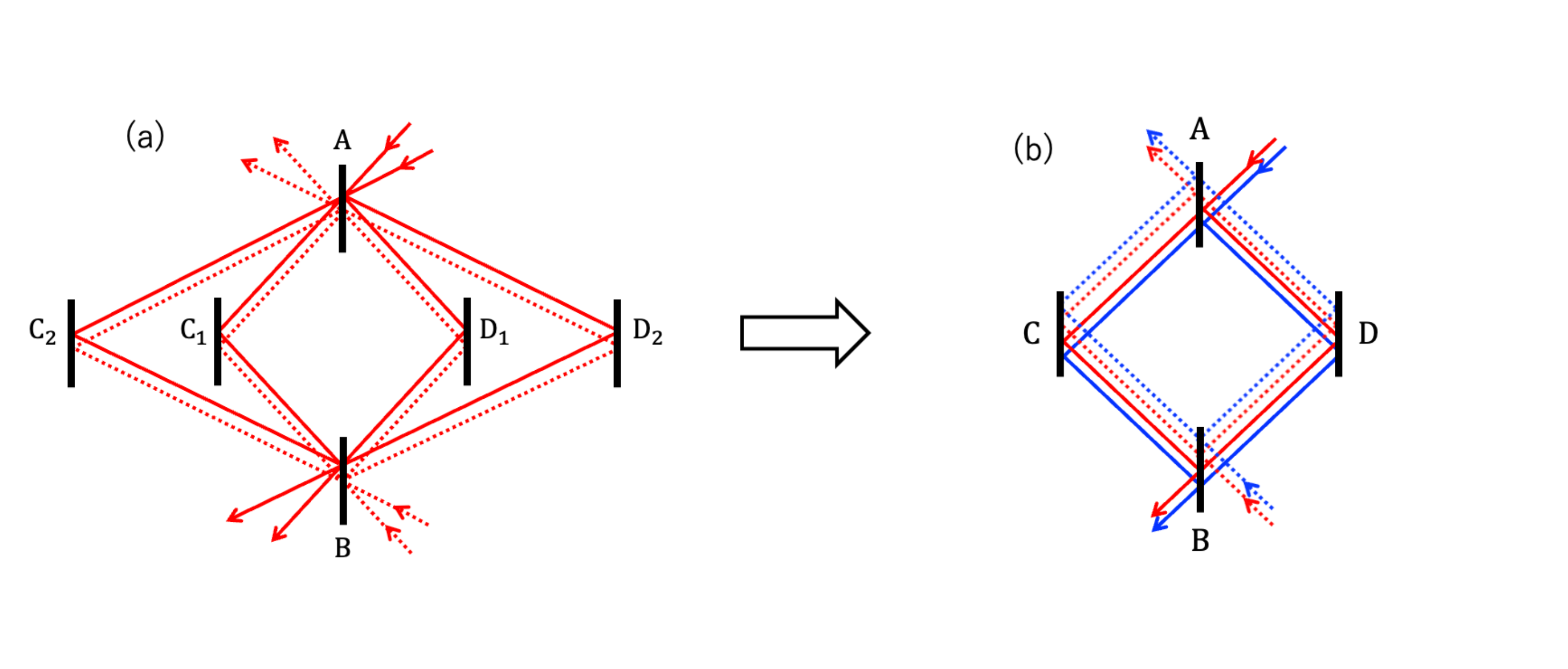}
\end{figure}

\newpage

\begin{figure}[h]
    \centering
    \caption{Configurations of the neutron DFI. Configuration (a) consists of two beamsplitters A and B and four mirrors $\mathrm{C}_1$, $\mathrm{C}_2$, $\mathrm{D}_1$, and $\mathrm{D}_2$, which constitute one large and one small MZIs. In configuration (a), two pairs of counter-propagating neutrons with the same speed enter one large and one small MZIs. Configuration (b) consists of two beamsplitters A and B and two mirrors C and D, which comprise a one MZI. In configuration (b), two bidirectional neutrons with slow (red) and fast (blue) velocities enter the single MZI. In this figure, the solid lines show the neutron trajectory incident from beamsplitter A and the dashed lines show the neutron trajectory incident from beamsplitter B.}
    \label{fig:1}
\end{figure}

\subsection{Cancellation of mirror displacement noise in the time domain}
\label{sec:2.2}

In configuration (b) of Figure \ref{fig:1}, two bidirectional neutrons with speeds $v_\mathrm{1}$ and $v_\mathrm{2}$ $(v_1>v_2)$ enter the single MZI. We consider the case where they hit each mirror at the same time $t=t'$. Fast and slow neutrons spend the times $T_\mathrm{1}$ and $T_\mathrm{2}$ $(T_1 < T_2)$, respectively, transmitting through one side of the MZI. Signals from GWs and other noise sources are registered as neutron phase shifts. Each phase shift from displacement noise is denoted by $\phi_{l_\mathrm{i}} (t)$ $(l = \mathrm{A, B, C, D} \ \mathrm{and} \ \mathrm{i} = 1,2)$, as shown in Table \ref{tab:1}.

The signal combination that cancels mirror displacement noise is defined as

\begin{align}
        & V_\mathrm{1}(t) = \phi_{\mathrm{{BA}_1}}(t)-\phi_{\mathrm{{AB}_1}}(t), \label{eq:1} \\
        & V_\mathrm{2}(t) = \phi_{\mathrm{{BA}_2}}(t)-\phi_{\mathrm{{AB}_2}}(t). \label{eq:2}
\end{align}

\noindent
The mirror displacement noise in these combinations can be canceled because each bidirectional neutron hits mirrors at the same time. Each neutron receives the same displacement noise at each point. As a result, the mirror displacement noise in a single MZI is canceled by the combinations of each bidirectional neutron in the time domain.

\begin{table}[h]
   \centering
   \caption{Phase shift resulting from displacement noise when four counter-propagating neutrons hit mirrors at the same time $t=t'$. The subscripts indicate the routes of the neutrons. For example, subscript BA ($\phi_\mathrm{{BA}_2}$) indicates the route $(\mathrm{B} \rightarrow \mathrm{C} \rightarrow \mathrm{A} \ \mathrm{and} \ \mathrm{B} \rightarrow \mathrm{D} \rightarrow \mathrm{A})$. The subscript 2 indicates slow and the subscript 1 indicates fast neutrons. \\}
   \label{tab:1}
   \scalebox{1}{
   {\renewcommand\arraystretch{2.0}
   \begin{tabular}{c||c|c|c|c} \hline
     Signal & C & D & B & A \\  \hline \hline
     $\phi_\mathrm{{BA}_2}(t)$ & $\phi_{\mathrm{C}_2} (t')$ & $\phi_{\mathrm{D}_2} (t')$ & $\phi_{\mathrm{B}_2} (t'-T_2)$ & $\phi_{\mathrm{A}_2} (t'+T_2)$ \\ \hline
     $\phi_\mathrm{{BA}_1}(t)$ & $\phi_{\mathrm{C}_1} (t')$ & $\phi_{\mathrm{D}_1} (t')$ & $\phi_{\mathrm{B}_1} (t'-T_1)$ & $\phi_{\mathrm{A}_1} (t'+T_1)$ \\ \hline
     $\phi_\mathrm{{AB}_2}(t)$ & $\phi_{\mathrm{C}_2} (t')$ & $\phi_{\mathrm{D}_2} (t')$ & $\phi_{\mathrm{B}_2} (t'+T_2)$ & $\phi_{\mathrm{A}_2} (t'-T_2)$ \\ \hline
     $\phi_\mathrm{{AB}_1}(t)$ & $\phi_{\mathrm{C}_1} (t')$ & $\phi_{\mathrm{D}_1} (t')$ & $\phi_{\mathrm{B}_1} (t'+T_1)$ & $\phi_{\mathrm{A}_1} (t'-T_1)$ \\  \hline
   \end{tabular}
   }
   }
\end{table}

\subsection{Cancellation of beam splitter displacement noise in the frequency domain}
\label{sec:2.3}

In Eq. (\ref{eq:1}) and (\ref{eq:2}), these combinations still have the displacement noise of the beamsplitters. This is because the neutrons do not hit the beamsplitters at the same time. With regard to displacement noise, a displacement can be represented as an exponential function. The amplitude and initial phase are given by $X_l$ and $\varphi_{l}$ $(l = \mathrm{A, B, C, D})$. With them, a beamsplitter displacement $x_l$ is given by

\begin{equation}
         x_l(t) = \sum_{\omega} X_{l}(\omega) e^{i \lparen \omega t + \varphi_{l}(\omega) \rparen}.
         \label{eq:3}
\end{equation}

\noindent
In the frequency domain of Eq. (\ref{eq:3}), an arbitrary term with an arbitrary frequency $\omega'$ is written as

\begin{equation}
        x_{l}(\omega') = X_{l}(\omega') e^{i \lparen \omega' t + \varphi_{l}(\omega') \rparen}.
        \label{eq:4}
\end{equation}

\noindent
Using the de Broglie wavelength, the phase shift $\phi_{l_\mathrm{i}}$ $(l = \mathrm{A, B, C, D} \ \mathrm{and} \ \mathrm{i} = 1,2)$  from displacement noise is given by
\begin{align}
         \phi_{l_\mathrm{i}} (\omega') &= \tfrac{2m}{\hbar} v_{\mathrm{i}} x_{l}(\omega')  \notag \\
         &= \tfrac{2m}{\hbar} v_{\mathrm{i}} X_{l}(\omega') e^{i \lparen \omega' t + \varphi_{l}(\omega') \rparen}.
         \label{eq:5}
\end{align}

\noindent
Defining $k_\mathrm{i}$ (i=1,2), a coefficient to simplify the equation, as
\begin{equation}
         k_\mathrm{i} = \tfrac{2m}{\hbar} v_{\mathrm{i}},
         \label{eq:6}
\end{equation}

\noindent
Eq. (\ref{eq:5}) can be written as

\begin{equation}
         \phi_{l_\mathrm{i}} (\omega') = X_{l}(\omega') k_\mathrm{i} e^{i \lparen \omega' t + \varphi_{l}(\omega') \rparen}.
         \label{eq:7}
\end{equation}

\noindent
For the signal combinations of $V_{\mathrm{1}}$ and $V_{\mathrm{2}}$, the phase shifts from the displacement of the beamsplitters are shown in Table \ref{tab:2}.

\begin{table}[htbp]
  \centering
  \scalebox{1}{
   {\renewcommand\arraystretch{2.0}
   \begin{tabular}{c||c|c}
     \hline
     Signal & B & A \\
     \hline \hline
     $V_{\mathrm{2}}$ & $ \phi_{B_\mathrm{2}}(\omega') \lbrace e^{ - i \omega' T_2 }- e^{ i \omega' T_2 } \rbrace $ & $ \phi_{A_\mathrm{2}}(\omega') \lbrace e^{ i \omega' T_2 }- e^{ - i \omega' T_2 } \rbrace $  \\ \hline
     $V_{\mathrm{1}}$ & $ \phi_{B_\mathrm{1}}(\omega') \lbrace e^{ - i \omega' T_1 }- e^{ i \omega' T_1 } \rbrace $ & $ \phi_{A_\mathrm{2}}(\omega') \lbrace e^{ i \omega' T_1 }- e^{ - i \omega' T_1 } \rbrace $  \\ \hline
   \end{tabular}
   }
   }
   \caption{Displacement noise of the beamsplitters in the Fourier domain when neutrons hit beamsplitters at the same time, $t=t'$.}
   \label{tab:2}
\end{table}

We define $\kappa_1$ and $\kappa_2$ as the coefficients that cancel displacement noise of the beamsplitters. With these coefficients, the combination $V_\mathrm{DFI}$ is given by

\begin{equation}
           V_\mathrm{DFI} = \kappa_1 V_1/k_1 - \kappa_2 V_2/k_2.
           \label{eq:8}
\end{equation}

\noindent
Here, we normalized $V_\mathrm{i}$ with $k_\mathrm{i}$. With regard to the displacement noise at beamsplitter B, coefficients $\kappa_1$ and $\kappa_2$ are given by

\begin{align}
        \kappa_1 V_1/k_1 - \kappa_2 V_2/k_2 & = X_{\mathrm{B}}(\omega') e^{\varphi_{\mathrm{B}}} \lbrack \kappa_1 \lbrace e^{ - i \omega' T_1 }- e^{ i \omega' T_1 } \rbrace - \kappa_2 \lbrace e^{ - i \omega' T_2 }- e^{ i \omega' T_2 }  \rbrace \rbrack  \notag \\
        & = X_{\mathrm{B}}(\omega') e^{\varphi_{\mathrm{B}}} \lbrack \kappa_1 \lbrace -2i \sin \omega' T_1 \rbrace - \kappa_2 \lbrace -2i \sin \omega' T_2 \rbrace \rbrack   \notag \\
        & = 0. \label{eq:9}
\end{align}
\begin{align}
        \therefore \ & \kappa_1 = \sin \omega' T_2 \label{eq:10}. \\
                     & \kappa_2 = \sin \omega' T_1 \label{eq:11}.
\end{align}

\noindent
Eq. (\ref{eq:8}) can be written as

\begin{equation}
           V_\mathrm{DFI} = \sin \omega' T_2 \cdot V_1/k_1 -  \sin \omega' T_1 \cdot V_2/k_2.
           \label{eq:12}
\end{equation}

\noindent
Although we discuss the displacement of beamsplitter B in Eq. (\ref{eq:9}), this combination can also cancel the displacement noise at beamsplitter A. In Eq. (\ref{eq:12}), the displacement noise of the beamsplitters can be canceled by the combination of different neutrons with frequency-dependent coefficients that are real. This cancellation is based on the symmetry condition for a neutron DFI. The neutron DFI configuration is symmetrical with respect to the mirrors, as the criteria. This symmetry means that the neutron propagation times between a beamsplitter and a mirror are also symmetrical. As a result, this condition of symmetry leads to cancellation with the frequency-dependent coefficients defined by the propagation time of the neutrons.

\subsection{Phasor diagram showing cancellation of beam splitter displacement noise}
\label{sec:2.4}

In this section, we show the mechanism for the cancellation of beamsplitter displacement noise using a phasor diagram. As shown in Eq. (\ref{eq:12}), the neutron DFI signal combination can cancel all the displacement noise in a single MZI. Provided that bidirectional neutrons hit each mirror at the same time, $t=0$, displacement noise at beamsplitters A and B are shown in Table \ref{tab:3}.

\begin{table}[htbp]
  \centering
  \scalebox{1}{
  {\renewcommand\arraystretch{2.0}
   \begin{tabular}{c||c|c}
     \hline
     Signal & B & A \\
     \hline \hline
     $V_{\mathrm{2}}$ & $ X_{\mathrm{B}} k_2 e^{\varphi_{\mathrm{B}}} \lbrace e^{ - i \omega \cdot T_2 }- e^{ i \omega \cdot T_2 } \rbrace $ &  $ X_{\mathrm{A}} k_2 e^{\varphi_{\mathrm{A}}} \lbrace e^{ i \omega \cdot T_2 }- e^{ - i \omega \cdot T_2 } \rbrace $ \\ \hline
     $V_{\mathrm{1}}$ & $ X_{\mathrm{B}} k_1 e^{\varphi_{\mathrm{B}}} \lbrace e^{ - i \omega \cdot T_1 }- e^{ i \omega \cdot T_1 } \rbrace $ &  $ X_{\mathrm{A}} k_1 e^{\varphi_{\mathrm{A}}} \lbrace e^{ i \omega \cdot T_1 }- e^{ - i \omega \cdot T_1 } \rbrace $ \\ \hline
   \end{tabular}
   }
   }
   \caption{Displacement noise of the beamsplitters when a neutron hits the beamsplitter at time $t=0$. $X_\mathrm{A}$ and $X_\mathrm{B}$ are the amplitudes of the position variations of beamsplitters, A and B. $\varphi_{A}$ and $\varphi_{B}$ are the initial phases of the beamsplitter position variations.}
   \label{tab:3}
\end{table}

\begin{figure}[h]
    \centering
    \includegraphics[clip,height=7.5cm]{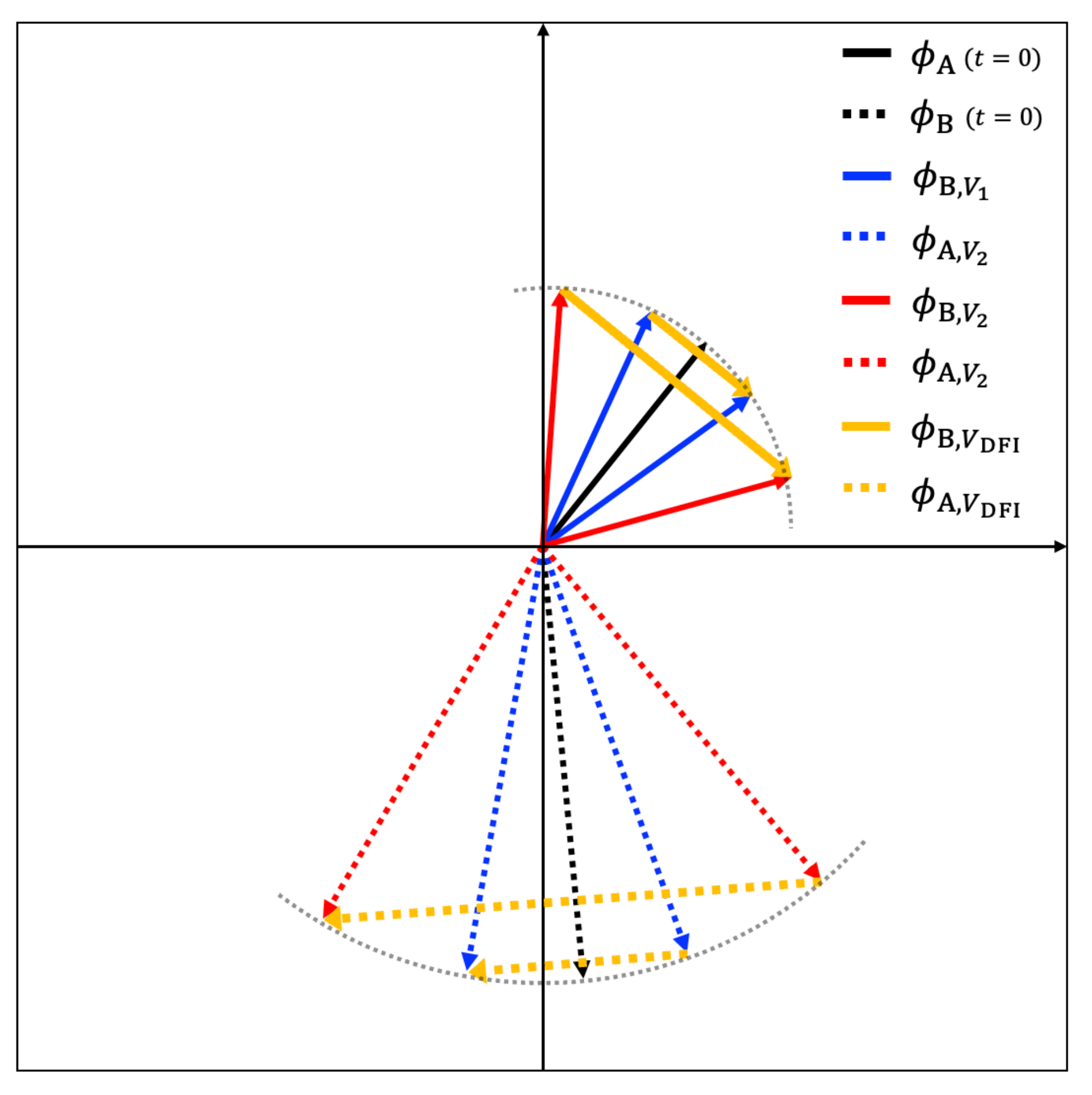}
    \caption{Phasor diagram of displacement noise at beamsplitters A and B in a neutron DFI. Arrow lengths show the noise amplitude. Angles of arrow rotation show the phase of the noise. Solid lines show the displacement noise of beamsplitter B. Dashed lines show the displacement noise of beamsplitter A. The displacement noise of the beamsplitters at $t=0$ are illustrated by black arrows, which are references for the blue and red arrows. The displacement noise, $\phi_{\mathrm{A},V_1}$ and $\phi_{\mathrm{B},V_1}$, when faster neutrons hit A and B, are illustrated by the blue arrows. The phase difference between the black and blue arrows is $\omega T_1$. The displacement noise, $\phi_{\mathrm{A},V_2}$ and $\phi_{\mathrm{B},V_2}$, when slower neutrons hit beamsplitters A and B are illustrated by the red arrows. The phase difference between the black and blue arrows is $\omega T_2$. The displacement noise in the DFI combination $V_\mathrm{DFI}$ are $\phi_{\mathrm{A},V_\mathrm{DFI}}$ and $\phi_{\mathrm{B},V_\mathrm{DFI}}$, which are illustrated by orange arrows.}
    \label{fig:2}
\end{figure}

This cancellation can be explained intuitively, using a phasor diagram. Figure \ref{fig:2} shows a phasor diagram of displacement noise at the beamsplitters when two bidirectional neutrons enter the MZI and hit the mirrors at $t=0$.

In Fig \ref{fig:2}, the $\phi_{\mathrm{B},V_\mathrm{DFI}}$ and $\phi_{\mathrm{A},V_\mathrm{DFI}}$ arrows are parallel, the difference between them is only their lengths. The arrow length indicates the amplitude of the noise. The parallelism of these arrows is attributed to the symmetry of the neutron DFI. In this condition, the ratio of the arrow lengths about beamsplitter B is equal to that about beamsplitter A. As shown in Eq. (\ref{eq:12}), the frequency-dependent coefficients $\kappa_1$ and $\kappa_2$ equalize the amplitudes of these arrows. Accordingly, the combination of a fast and a slow neutron can cancel beamsplitter displacement noise in a neutron DFI.

\subsection{Gravitational effect on neutron trajectory}
\label{sec:2.5}

\begin{figure}[h]
    \centering
    \includegraphics[clip,height=10cm]{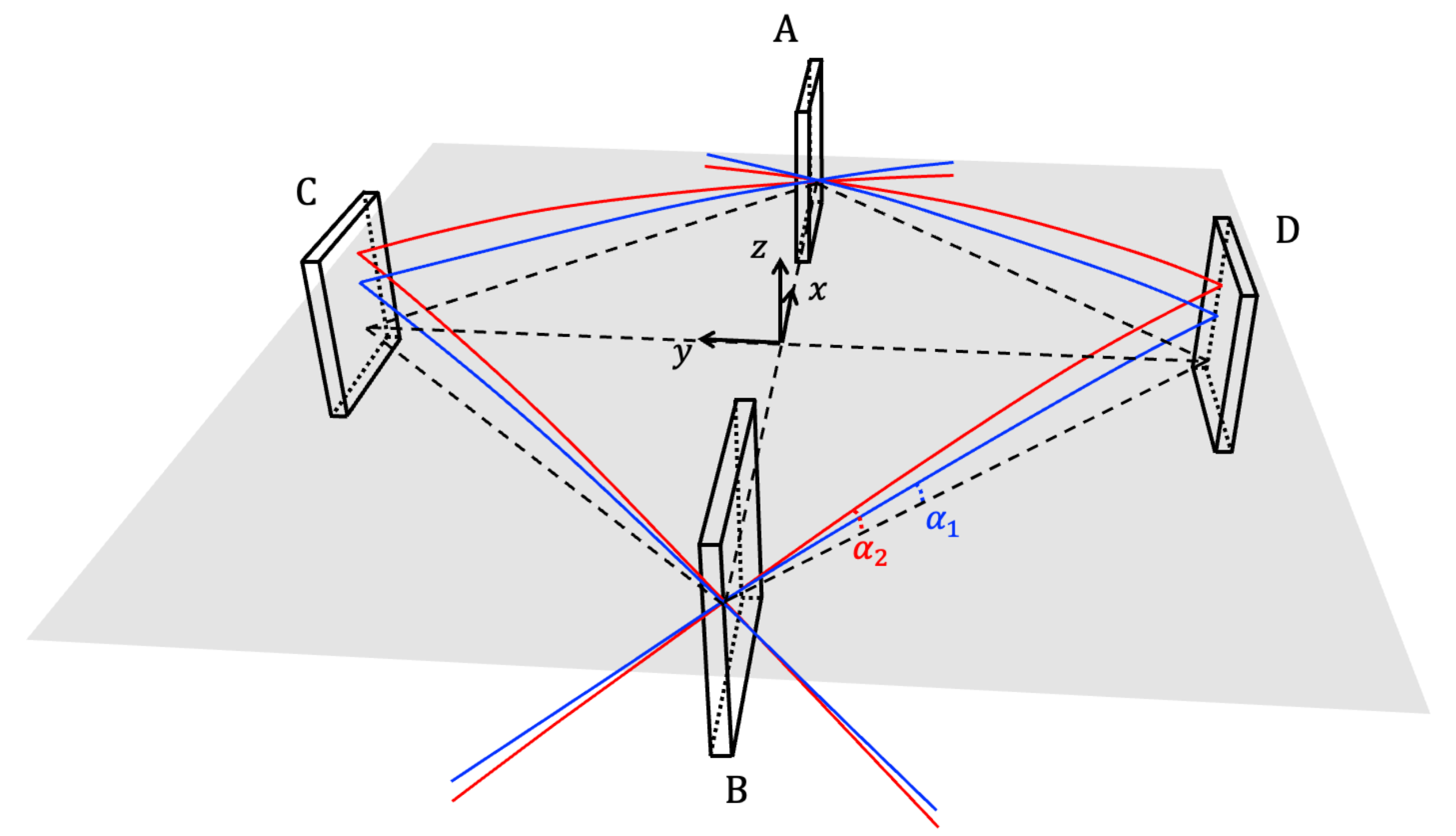}
    \caption{Neutron trajectories in three dimensions. The dashed lines connect A, B, C, and D on the $x$-$y$ plane. The trajectory of each neutron is shown by the blue (fast neutron) and red (slow neutron) lines. The initial angles between the neutron trajectories and the $x$-$y$ plane are $\alpha_1$ and $\alpha_2$.}
    \label{fig:3}
\end{figure}

A neutron trajectory in a single MZI is parabolic because of gravity. In Figure \ref{fig:3}, the neutron DFI configuration is shown in three dimensions. The angles between the neutron trajectories and the $x$-$y$ plane are $\alpha_1$ and $\alpha_2$ for the neutrons with the fast and slow speeds, respectively. Under the condition that all neutrons hit the same points on beamsplitters A and B at $z=0$, the angles $\alpha_1$ and $\alpha_2$ are constrained by

\begin{equation}
         \sin 2 \alpha_\mathrm{i} = \frac{2 g L}{{v_{\mathrm{i}}^2}}.
         \label{eq:13}
\end{equation}

In Fig \ref{fig:3}, the fast and slow neutrons hit different points on the mirrors, although they hit the same points on the beamsplitters. Mirror noise cancellation is possible because of the combination of signals from the bidirectional neutrons.

\newpage

\section{Response of a neutron DFI to gravitational waves}

\label{sec:3}

\begin{figure}[h]
    \centering
    \includegraphics[clip,height=8cm]{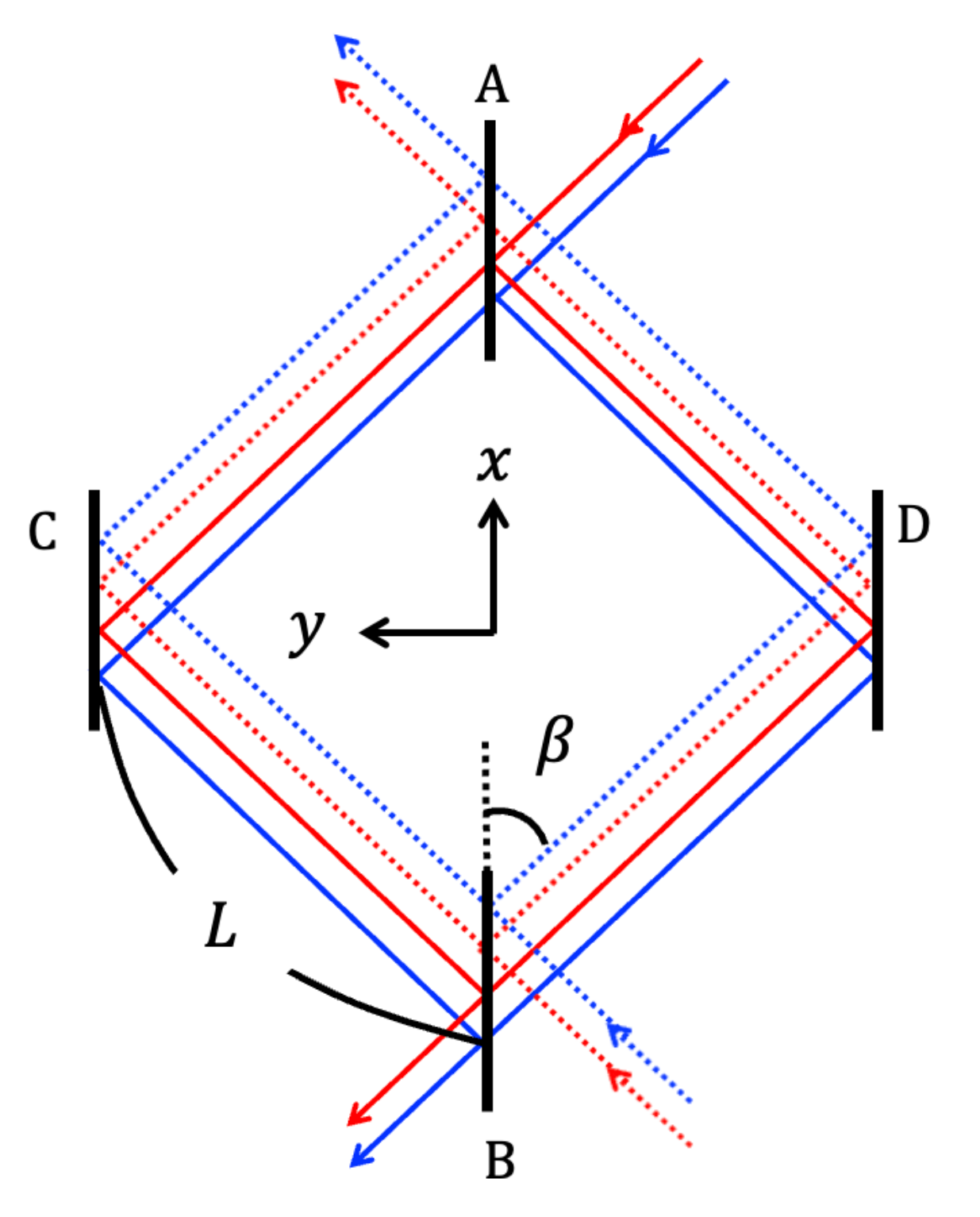}
    \caption{Definition of parameters in the neutron DFI configuration.}
    \label{fig:4}
\end{figure}

The neutron DFI configuration and parameter definitions are shown in Fig \ref{fig:4}. The time each neutron spends transiting between the beamsplitters and the mirrors is $T_\mathrm{i}$, which is given by

\begin{equation}
         T_\mathrm{i} = \frac{L}{v_{\mathrm{i}} \cos \alpha_\mathrm{i}}.
         \label{eq:14}
\end{equation}

\noindent
The coordinates of the neutron incidence point on beamsplitter B are given by

\begin{equation}
      \mathbf{x}_\mathrm{B} = \lbrace  -L \cos \beta , 0 ,0  \rbrace.
      \label{eq:15}
\end{equation}

\noindent
The neutron trajectory (B→C) is given by

\begin{equation}
         \mathbf{x}_\mathrm{{BC}_i} (t) = \lbrace v_\mathrm{i} t \cos \alpha_\mathrm{i} \cos \beta , v_\mathrm{i} t \cos \alpha_\mathrm{i} \sin \beta , v_\mathrm{i} t \sin \alpha_\mathrm{i} - \frac{g}{2} t^2 \rbrace \ \ \ \ \ (0 \leq t \leq T_\mathrm{i}),
         \label{eq:16}
\end{equation}
\normalsize

\noindent
and the coordinates of the impact points on mirror C for each neutron are given by
\begin{align}
      \mathbf{x}_\mathrm{C_i} &= \mathbf{x}_\mathrm{B} + \mathbf{x}_\mathrm{{BC}_i} (T_\mathrm{i}) \notag \\
                     &= \lbrace 0 , L \sin \beta , \frac{g}{2} T_\mathrm{i}^2 \rbrace.
                     \label{eq:17}
\end{align}

\noindent
The neutron trajectory (C→A) is given by

\begin{equation}
         \mathbf{x}_\mathrm{{CA}_i} (t) = \lbrace v_\mathrm{i} t \cos \alpha_\mathrm{i} \cos \beta , - v_\mathrm{i} t \cos \alpha_\mathrm{i} \sin \beta , - \frac{g}{2} t^2 \rbrace \ \ \ \ \ (0 \leq t \leq T_\mathrm{i}),
         \label{eq:18}
\end{equation}

\noindent
and the coordinate of the incidence points on the beamsplitter are given by
\begin{align}
      \mathbf{x}_\mathrm{A_i} &= \mathbf{x}_\mathrm{C} + \mathbf{x}_\mathrm{{CA}_i} (T_\mathrm{i}) \notag \\
                     &= \lbrace  L \cos \beta, 0 ,0  \rbrace.
                     \label{eq:19}
\end{align}

\noindent
The neutron trajectory (B→D) is given by

\begin{equation}
         \mathbf{x}_\mathrm{{BD}_i} (t) = \lbrace v_\mathrm{i} t \cos \alpha_\mathrm{i} \cos \beta , - v_\mathrm{i} t \cos \alpha_\mathrm{i} \sin \beta , v_\mathrm{i} t \sin \alpha_\mathrm{i} - \frac{g}{2} t^2 \rbrace \ \ \ \ \ (0 \leq t \leq T_\mathrm{i}),
         \label{eq:20}
\end{equation}
\normalsize

\noindent
and the coordinates of the impact points on mirror D for each neutron are given by
\begin{align}
      \mathbf{x}_\mathrm{D_i} &= \mathbf{x}_\mathrm{B} + \mathbf{x}_\mathrm{{BD}_i} (T_\mathrm{i}) \notag \\
                     &= \lbrace 0 , - L \sin \beta , v_\mathrm{i} T_\mathrm{i} \sin \alpha_\mathrm{i} - \frac{g}{2} T_\mathrm{i}^2 \rbrace.
                     \label{eq:21}
\end{align}

\noindent
The neutron trajectory (D→A) is given by

\begin{equation}
         \mathbf{x}_\mathrm{{DA}_i} (t) = \lbrace v_\mathrm{i} t \cos \alpha_\mathrm{i} \cos \beta , v_\mathrm{i} t \cos \alpha_\mathrm{i} \sin \beta , - \frac{g}{2} t^2 \rbrace \ \ \ \ \ (0 \leq t \leq T_\mathrm{i}).
         \label{eq:22}
\end{equation}

\noindent
The wavenumbers of the neutrons propagating from B to A are given by


\begin{align}
        & \mathbf{k}_\mathrm{{BC}_i} (t) = \frac{m}{\hbar} \lbrace v_\mathrm{i} \cos \alpha_\mathrm{i} \cos \beta , v_\mathrm{i} \cos \alpha_\mathrm{i} \sin \beta , v_\mathrm{i} \sin \alpha_\mathrm{i} - g t \rbrace, \label{eq:23}\\
        & \mathbf{k}_\mathrm{{CA}_i} (t) = \frac{m}{\hbar} \lbrace v_\mathrm{i} \cos \alpha_\mathrm{i} \cos \beta , - v_\mathrm{i} \cos \alpha_\mathrm{i} \sin \beta , - g t \rbrace, \label{eq:24}\\
        & \mathbf{k}_\mathrm{{BD}_i} (t) = \frac{m}{\hbar} \lbrace v_\mathrm{i} \cos \alpha_\mathrm{i} \cos \beta , - v_\mathrm{i} \cos \alpha_\mathrm{i} \sin \beta , v_\mathrm{i} \sin \alpha_\mathrm{i} - g t \rbrace, \label{eq:25}\\
        & \mathbf{k}_\mathrm{{DA}_i} (t) = \frac{m}{\hbar} \lbrace v_\mathrm{i} \cos \alpha_\mathrm{i} \cos \beta , v_\mathrm{i} \cos \alpha_\mathrm{i} \sin \beta , - g t \rbrace \ \ \ (0 \leq t \leq T_\mathrm{i}). \label{eq:26}
\end{align}

\normalsize

\noindent
In the same way, the wavenumbers of the neutrons propagating from A to B are given by


\begin{align}
        & \mathbf{k}_\mathrm{{AC}_i} (t) = \frac{m}{\hbar} \lbrace - v_\mathrm{i} \cos \alpha_\mathrm{i} \cos \beta , v_\mathrm{i} \cos \alpha_\mathrm{i} \sin \beta , v_\mathrm{i} \sin \alpha_\mathrm{i} - g t \rbrace, \label{eq:27}\\
        & \mathbf{k}_\mathrm{{CB}_i} (t) = \frac{m}{\hbar} \lbrace - v_\mathrm{i} \cos \alpha_\mathrm{i} \cos \beta , - v_\mathrm{i} \cos \alpha_\mathrm{i} \sin \beta , - g t \rbrace, \label{eq:28}\\
        & \mathbf{k}_\mathrm{{AD}_i} (t) = \frac{m}{\hbar} \lbrace - v_\mathrm{i} \cos \alpha_\mathrm{i} \cos \beta , - v_\mathrm{i} \cos \alpha_\mathrm{i} \sin \beta , v_\mathrm{i} \sin \alpha_\mathrm{i} - g t \rbrace, \label{eq:29}\\
        & \mathbf{k}_\mathrm{{DB}_i} (t) = \frac{m}{\hbar} \lbrace - v_\mathrm{i} \cos \alpha_\mathrm{i} \cos \beta , v_\mathrm{i} \cos \alpha_\mathrm{i} \sin \beta , - g t \rbrace \ \ \ (0 \leq t \leq T_\mathrm{i}). \label{eq:30}
\end{align}

\normalsize

\noindent
For the first half of the parabolic trajectory, the initial speed of the neutrons is $v_\mathrm{i}$. On the other hand, for the second half of the parabolic trajectory, the initial speed the of neutrons is $v_\mathrm{i} \cos \alpha_\mathrm{i}$. The normalized wavenumbers of the neutrons are given by

\begin{equation}
       \tilde{\mathbf{k}}_\mathrm{{BC}_i} (t) = \mathbf{k}_\mathrm{{BC}_i} / v_\mathrm{i}, \ \ \  \tilde{\mathbf{k}}_\mathrm{{BD}_i} (t) = \mathbf{k}_\mathrm{{BD}_i} / v_\mathrm{i}, \ \ \         \tilde{\mathbf{k}}_\mathrm{{AC}_i} (t) = \mathbf{k}_\mathrm{{AC}_i} / v_\mathrm{i}, \ \ \   \tilde{\mathbf{k}}_\mathrm{{AD}_i} (t) = \mathbf{k}_\mathrm{{AD}_i} / v_\mathrm{i}, \label{eq:31}
\end{equation}

\begin{equation}
       \tilde{\mathbf{k}}_\mathrm{{CA}_i} (t) = \mathbf{k}_\mathrm{{CA}_i} / v_\mathrm{i} \cos \alpha_\mathrm{i}, \ \ \ \tilde{\mathbf{k}}_\mathrm{{DA}_i} (t) = \mathbf{k}_\mathrm{{DA}_i} / v_\mathrm{i} \cos \alpha_\mathrm{i}, \ \ \ \tilde{\mathbf{k}}_\mathrm{{CB}_i} (t) = \mathbf{k}_\mathrm{{CB}_i} / v_\mathrm{i} \cos \alpha_\mathrm{i}, \ \ \ \tilde{\mathbf{k}}_\mathrm{{DB}_i} (t) = \mathbf{k}_\mathrm{{DB}_i} / v_\mathrm{i} \cos \alpha_\mathrm{i}. \label{eq:32}
\end{equation}

\noindent
The phase shift of the neutrons caused by GWs is $\phi_\mathrm{gw}$, which is derived from the Klein-Gordon equation \cite{DFNI_Nishizawa}. The phase shift $\phi_\mathrm{gw}$ is defined by

\begin{equation}
         \frac{\partial \phi_\mathrm{gw}}{\partial t} \approx - \frac{h^{ij} k_i k_j}{2 m}.
         \label{eq:33}
\end{equation}

\noindent
In this neutron DFI configuration, when a fast neutron propagates from B to A, the phase shift due to GWs is given by

\begin{equation}
         \phi^\mathrm{gw}_\mathrm{{BC}_\mathrm{i}} (t) = -\frac{v_\mathrm{i}^2}{2 m} \int_{t}^{t+T_\mathrm{i}} h^{ij} \lbrack t'-t , \mathbf{x}_\mathrm{BC_i} (t'-t) \rbrack  \tilde{k}_{\mathrm{{BC}_i}i} (t'-t) \tilde{k}_{\mathrm{{BC}_i}j} (t'-t) dt'.
         \label{eq:34}
\end{equation}
\normalsize

\noindent
We define the timing noise $\phi^\mathrm{clock}_\mathrm{{BC}_1} (t)$ and the displacement noise $\phi^\mathrm{disp}_\mathrm{{BC}_1} (t)$ as

\begin{equation}
       \phi^\mathrm{clock}_\mathrm{{BC}_\mathrm{i}} (t) \approx m \lbrace \tau_\mathrm{C} (t+T_\mathrm{i}) - \tau_\mathrm{B} (t) \rbrace,
       \label{eq:35}
\end{equation}

\begin{equation}
       \phi^\mathrm{disp}_\mathrm{{BC}_\mathrm{i}} (t) = \phi_{\mathrm{C}_\mathrm{i}} (t+T_\mathrm{i}) - \phi_{\mathrm{B}_\mathrm{i}} (t).
       \label{eq:36}
\end{equation}

\noindent
Here, $\tau_l$ is the clock noise at location $l$ $(l = \mathrm{A, B, C, D})$. At the GW angular frequency $\Omega$, the Fourier transform of $h^{ij}$ is defined as

\begin{equation}
         H^{ij} (\Omega) \equiv \int_{- \infty}^{\infty} dt e^{i \Omega t} h^{ij} \lbrack t' , \mathbf{x}_\mathrm{i} (t') \rbrack.
         \label{eq:37}
\end{equation}

\noindent
In the Fourier domain, the GW signal $\Phi^\mathrm{gw}_\mathrm{{BC}_i} (\Omega)$ is given by


\begin{align}
         \Phi^\mathrm{gw}_\mathrm{BC_i} (\Omega) &= - \frac{v_\mathrm{i}^2}{2 m} \left \lbrace P_0 (\Omega) \tilde{k}_{\mathrm{BC_i}I} \tilde{k}_{\mathrm{BC_i}J} H^{IJ} (\Omega) \notag \right. \\
         & \left. + \frac{m}{\hbar} \left ( \sin \alpha_\mathrm{i} P_0 (\Omega) + \frac{g}{v_\mathrm{i}} P_1 (\Omega) \right ) \tilde{k}_{\mathrm{BC_i}I} H^{Iz} (\Omega) \notag \right. \\
         & \left. + \left(\frac{m}{\hbar} \right)^2 \left (\sin^2 \alpha_\mathrm{i} P_0 (\Omega) + 2 \frac{g}{v_\mathrm{i}} \sin \alpha_\mathrm{i} P_1 (\Omega) + \frac{g^2}{v_\mathrm{i}^2} P_2 (\Omega) \right )H^{zz} (\Omega) \right \rbrace,
         \label{eq:38}
\end{align}

\normalsize

\noindent
and the clock noise $\Phi^\mathrm{clock}_\mathrm{{BC}_\mathrm{i}} (\Omega)$ and the detector noise $\Phi^\mathrm{disp}_\mathrm{{BC}_\mathrm{i}} (\Omega)$ are given by

\begin{equation}
       \Phi^\mathrm{clock}_\mathrm{{BC}_\mathrm{i}} (\Omega) \approx \omega_\mathrm{i} (\Omega) m \lbrace \tau_\mathrm{C} (\Omega) - \tau_\mathrm{B} (\Omega) \rbrace \ \mathrm{and}
       \label{eq:39}
\end{equation}

\begin{equation}
       \Phi^\mathrm{disp}_\mathrm{{BC}_\mathrm{i}} (\Omega) = \omega_\mathrm{i} (\Omega)  \phi_{\mathrm{C}_\mathrm{i}} (\Omega) - \phi_{\mathrm{B}_\mathrm{i}} (\Omega).
       \label{eq:40}
\end{equation}

\noindent
We define the following parameters

\begin{equation}
         \omega_\mathrm{i} (\Omega) \equiv e^{- i \Omega T_\mathrm{i}},
         \label{eq:41}
\end{equation}

\begin{equation}
         P_0 (\Omega) \equiv -\frac{i}{\Omega} \lbrace 1 - \omega_\mathrm{i} (\Omega) \rbrace,
         \label{eq:42}
\end{equation}

\begin{equation}
         P_1 (\Omega) \equiv \frac{1}{\Omega^2} \lbrace 1 - \omega_\mathrm{i} (\Omega)  (1 + i \Omega T_\mathrm{i} ) \rbrace,
         \label{eq:43}
\end{equation}

\begin{equation}
         P_2 (\Omega) \equiv \frac{2i}{\Omega^3} \left \lbrace 1 - \omega_\mathrm{i} (\Omega)  \left(1 + i \Omega T_\mathrm{i} - \frac{1}{2} \Omega^2 T_\mathrm{i}^2 \right) \right \rbrace.
         \label{eq:44}
\end{equation}

\noindent
With Eq. (\ref{eq:38})-(\ref{eq:40}), the signal resulting from propagation from B to C is given in the Fourier domain by


\begin{align}
         \Phi_\mathrm{BC_i} (\Omega) &= - \frac{v_\mathrm{i}^2}{2 m} \left \lbrace P_0 (\Omega) \tilde{k}_{\mathrm{BC_i}I} \tilde{k}_{\mathrm{BC_i}J} H^{IJ} (\Omega) \notag \right. \\
         & \left. + \frac{m}{\hbar} \left ( \sin \alpha_\mathrm{i} P_0 (\Omega) + \frac{g}{v_\mathrm{i}} P_1 (\Omega) \right ) \tilde{k}_{\mathrm{BC_i}I} H^{Iz} (\Omega) \notag \right. \\
         & \left. + \left(\frac{m}{\hbar} \right)^2 \left (\sin^2 \alpha_\mathrm{i} P_0 (\Omega) + 2 \frac{g}{v_\mathrm{i}} \sin \alpha_\mathrm{i} P_1 (\Omega) + \frac{g^2}{v_\mathrm{i}^2} P_2 (\Omega) \right )H^{zz} (\Omega) \right \rbrace, \notag \\
         & + \Phi^\mathrm{clock}_\mathrm{{BC}_\mathrm{i}} (\Omega) + \Phi^\mathrm{disp}_\mathrm{{BC}_\mathrm{i}} (\Omega).
         \label{eq:45}
\end{align}

\normalsize

\noindent
The GW response function is given by


\begin{align}
       R_{\mathrm{BC_i}} (\Omega) &= \frac{1}{|H|} \left(\frac{\hbar}{m} \right)^2 \Bigl| P_0 (\Omega) \tilde{k}_{\mathrm{{BC}_\mathrm{i}}I} \tilde{k}_{\mathrm{{BC}_\mathrm{i}}J} H^{IJ} (\Omega) \Bigr. \notag \\
       & \Bigl. + \frac{m}{\hbar}  \left( \sin \alpha_\mathrm{i} P_0 (\Omega) + \frac{g}{v_\mathrm{i}} P_1 (\Omega) \right) \tilde{k}_{\mathrm{{BC}_\mathrm{i}}I} H^{Iz} (\Omega) \Bigr. \notag \\
       & \Bigl. + \left(\frac{m}{\hbar} \right)^2 \left (\sin^2 \alpha_\mathrm{i} P_0 (\Omega) + 2 \frac{g}{{v_\mathrm{i}}} \sin \alpha_\mathrm{i} P_1 (\Omega) + \frac{g^2}{{v_\mathrm{i}}^2} P_2 (\Omega)\right) H^{zz} (\Omega) \Bigr | .
       \label{eq:46}
\end{align}

\normalsize

\noindent
Signals for the detector shown in Figure \ref{fig:4} are given by

\begin{align}
        \phi_{\mathrm{BA}_\mathrm{i}} (t) = & \phi_{\mathrm{BC_i}} (t) + \phi_{\mathrm{CA_i}} (t+T_\mathrm{i}) \notag \\
        & -\phi_{\mathrm{BD_i}} (t) - \phi_{\mathrm{DA_i}} (t+T_\mathrm{i}),
        \label{eq:47}
\end{align}

\begin{align}
        \phi_{\mathrm{AB}_\mathrm{i}} (t) = & \phi_{\mathrm{AC_i}} (t) + \phi_{\mathrm{CB_i}} (t+T_\mathrm{i}) \notag \\
        & -\phi_{\mathrm{AD_i}} (t) - \phi_{\mathrm{DB_i}} (t+T_\mathrm{i}).
        \label{eq:48}
\end{align}

\noindent
In the Fourier domain, these signals are given by

\begin{align}
        \Phi_{\mathrm{BA}_\mathrm{i}} (\Omega) = & \Phi_{\mathrm{BC_i}} (\Omega) + \omega_\mathrm{i} (\Omega) \Phi_{\mathrm{CA_i}} (\Omega) \notag \\
        & -\Phi_{\mathrm{BD_i}} (\Omega) - \omega_\mathrm{i} (\Omega)  \Phi_{\mathrm{DA_i}} (\Omega),
        \label{eq:49}
\end{align}

\begin{align}
        \Phi_{\mathrm{AB}_\mathrm{i}} (\Omega) = & \Phi_{\mathrm{AC_i}} (\Omega) + \omega_\mathrm{i} (\Omega) \Phi_{\mathrm{CB_i}} (\Omega) \notag \\
        & -\Phi_{\mathrm{AD_i}} (\Omega) - \omega_\mathrm{i} (\Omega)  \Phi_{\mathrm{DB_i}} (\Omega).
        \label{eq:50}
\end{align}

\noindent
From Eq. (\ref{eq:1})-(\ref{eq:2}), the signal combination that cancels mirror displacement noise is given by

\begin{equation}
        \Phi_{V_\mathrm{i}} (\Omega) = \Phi_{\mathrm{BA}_\mathrm{i}} (\Omega) - \Phi_{\mathrm{AB}_\mathrm{i}} (\Omega).
        \label{eq:51}
\end{equation}

\noindent

Accordingly, the neutron DFI signal in the Fourier domain is given by

\begin{equation}
        \Phi_\mathrm{DFI} (\Omega) = \frac{1}{2 \Omega \bar{T}} \left \lbrace \gamma_1 (\Omega) \Phi_{V_\mathrm{1}} (\Omega) -  \gamma_2 (\Omega) \Phi_{V_\mathrm{2}} (\Omega) \right \rbrace,
        \label{eq:52}
\end{equation}

\begin{equation}
        \gamma_1 (\Omega) = \frac{\sin \Omega T_2}{v_1 \cos\alpha_\mathrm{1} \sin\beta}, \ \ \ \  \gamma_2 (\Omega) = \frac{\sin \Omega T_1}{v_2 \cos\alpha_\mathrm{2} \sin\beta}.
        \label{eq:53}
\end{equation}

\noindent

Here, $\gamma_\mathrm{i} (\Omega)$ contains the frequency-dependent coefficients required to cancel the displacement noise and normalization terms. The division by $2 \Omega \bar{T}$ plays a role in maintaining the neutron DFI response at lower frequencies. When a GW with a strain $h_{ij}$ and a polarization angle $\psi$ propagates from an arbitrary direction $(\phi,\theta)$, the rotation matrix is given by


\begin{equation}
     \mathcal{R} =
     \begin{pmatrix}
       \cos \phi & \sin \phi & 0 \\
       - \sin \phi & \cos \phi & 0 \\
       0 & 0 & 1
     \end{pmatrix}
     \begin{pmatrix}
       \cos \theta & 0 & - \sin \theta \\
       0 & 1 & 0 \\
       \sin \theta & 0 & \cos \theta
     \end{pmatrix}
     \begin{pmatrix}
       \cos \psi & \sin \psi & 0 \\
       - \sin \psi & \cos \psi & 0 \\
       0 & 0 & 1
     \end{pmatrix},
     \label{eq:54}
\end{equation}

\normalsize

\noindent
and the GW strain $h'_{ij}$ is written as

\begin{equation}
     h'_{ij} = \mathcal{R}_{ia} \mathcal{R}_{ib} h_{ab} = (\mathcal{R} h \mathcal{R}^T)_{ij}.
     \label{eq:55}
\end{equation}

\begin{figure}[htbp]
  \begin{tabular}{cc}
     \begin{minipage}[b]{0.45\hsize}
       \centering
       \includegraphics[clip,width=7.5cm]{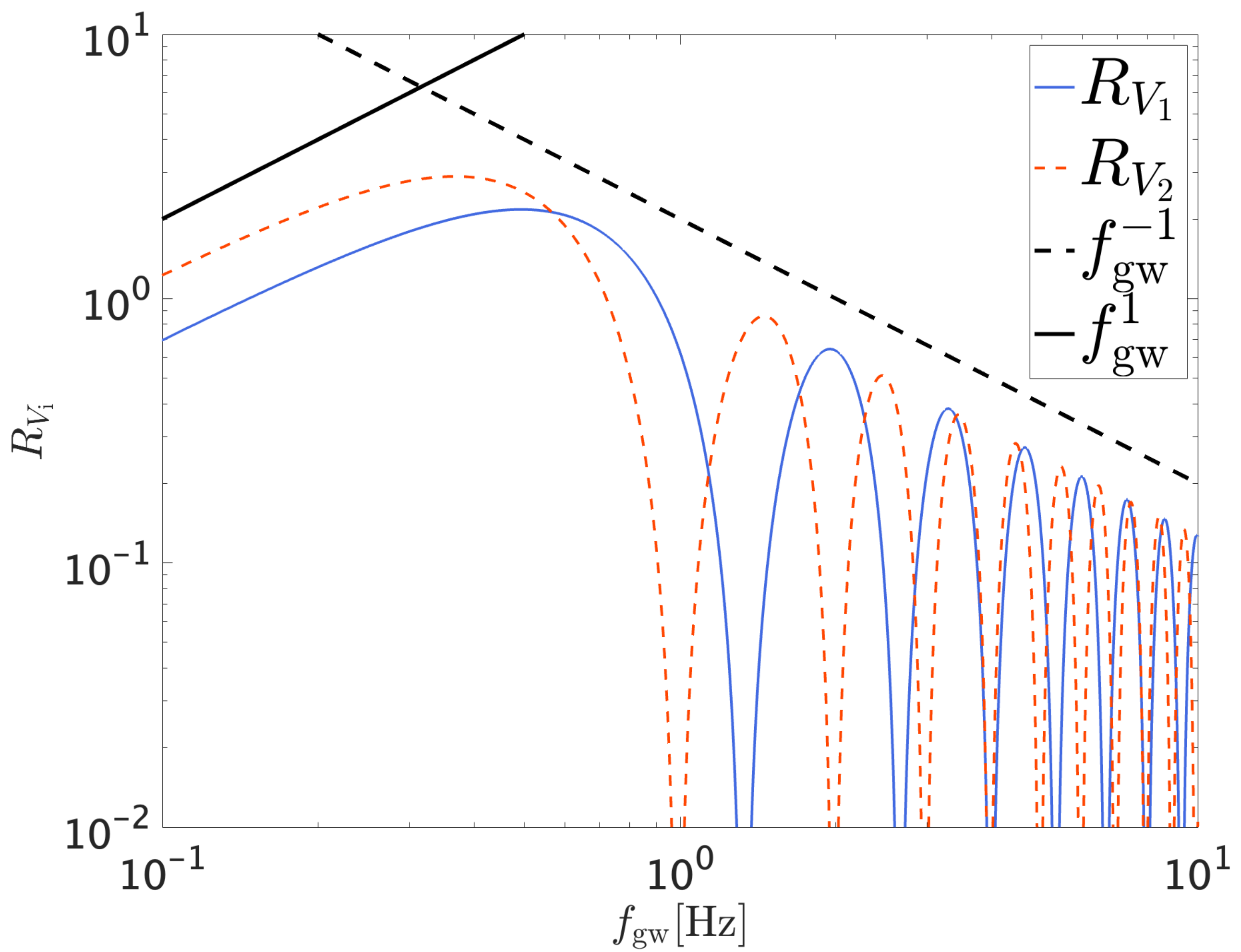}
     \end{minipage} &
     \begin{minipage}[b]{0.45\hsize}
       \centering
       \includegraphics[clip,width=7.5cm]{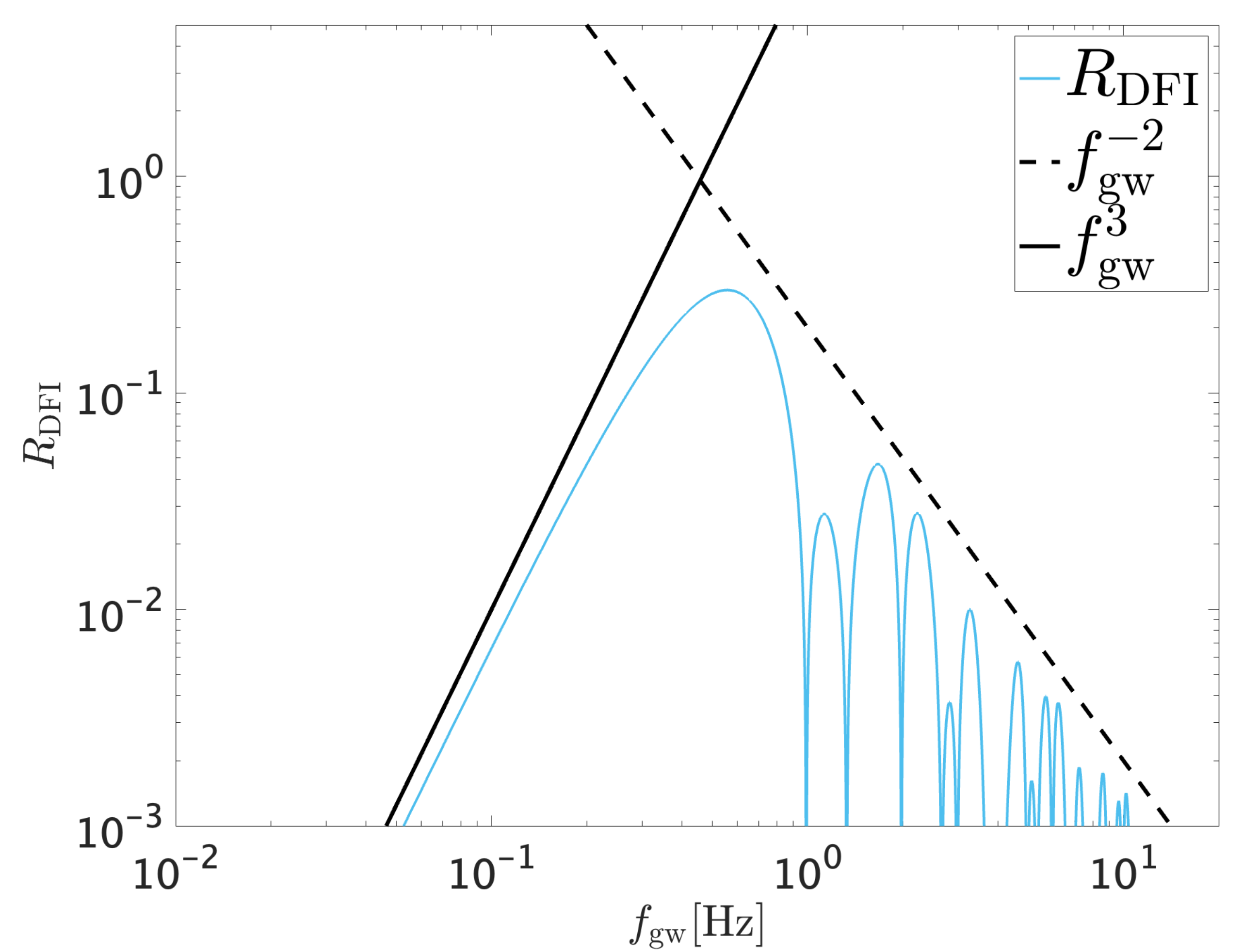}
     \end{minipage}
  \end{tabular}
    \caption{Response to GWs with the polarization of the cross mode ($\psi=\pi/4$). The left panel shows the response of a single MZI with a fast (blue solid curve) and a slow (red dashed curve) bidirectional neutron. In the left panel, the solid and dashed black lines are proportional to $f_\mathrm{gw}^1$ and $f_\mathrm{gw}^{-1}$. The right panel shows the neutron DFI response of a single MZI with two bidirectional neutrons. The solid and dashed black lines are proportional to $f_\mathrm{gw}^3$ and $f_\mathrm{gw}^{-2}$.}
    \label{fig:5}
\end{figure}

When a GW with the polarization of the cross mode ($\psi=\pi/4$) propagates along the z axis $(\theta=0,\phi=0)$, the response to GWs in a single MZI with two different-speed bidirectional neutrons is shown in Figure \ref{fig:5}. For $L=75$ m, $v_{0,1}=100$ m/s, $v_{0,2}=75$ m/s, and $\beta=\pi/4$ rad, the other parameters derived from Eq.(\ref{eq:13}) and (\ref{eq:14}) are $T_1=0.75$ s, $T_2=1.00$ s, $\alpha_1=4.23$ deg, and $\alpha_2=7.57$ deg.

The single-MZI signal of the fast or slow bidirectional neutrons, which is given by Eq.(\ref{eq:51}), cancels only mirror displacement noise. In the left panel of Figure \ref{fig:5}, the GW response in each combination is proportional to $f_\mathrm{gw}^1$ at lower frequencies. The peak of each response curve is located around 1Hz and these curves are proportional to $f_\mathrm{gw}^{-1}$ at higher frequencies. This GW response of the signal combination with fast or slow bidirectional neutrons has dips at frequencies determined by ${T_1}^{-1}$ or ${T_2}^{-1}$.

The DFI signal that combines the two single MZI signals of the fast and slow bidirectional neutrons, which is shown in Eq.(\ref{eq:52}), cancels all displacement noise of the mirrors and beamsplitters. In the right panel of Figure \ref{fig:5}, the GW response in the neutron DFI combination is proportional to $f_\mathrm{gw}^3$ at lower frequencies and has a peak around 0.6 Hz. The neutron DFI response at higher frequencies is proportional to $f_\mathrm{gw}^{-2}$ and has dips in the same manner as the combination of the $V_1$ and $V_2$ curves, which are shown in the left panel of Figure \ref{fig:5}.

It should be noted that the GW response function in this configuration is similar to that in the two-MZI configuration (large and small MZIs) in the previous research \cite{DFNI_Nishizawa}. The attainable sensitivity and technical challenges with this configuration are also similar to those with the two-MZI configuration, which are discussed in \cite{DFNI_Nishizawa}.

\section{Conclusions}

In this research, we have simplified the neutron DFI configuration by replacing bidirectional neutrons with the same speed in two MZIs with bidirectional neutrons with different speeds in a single MZI. This simplification is possible because the speed of a neutron can be changed arbitrarily, which is not possible with laser light. In the time domain, mirror displacement noise can be canceled when the bidirectional neutrons hit the mirrors at the same time. In the frequency domain, beamsplitter displacement noise can be canceled with the frequency-dependent coefficients defined by the propagation time of the neutrons. This cancellation is based on the condition that the neutron DFI has a configuration that is symmetrical with respect to orientation in which bidirectional neutrons hit the mirrors at the same time. This cancellation can be explained visually in a phasor diagram, which makes it possible to understand the noise cancellation mechanism intuitively. This simplification of the neutron DFI configuration will increase the possibility of detecting primordial GWs by a neutron DFI in the future.

\section*{Acknowledgement}

We would like to thank Rick Savage for English editing. This work was supported by the Japan Society for the Promotion of Science (JSPS) KAKENHI Grant Number JP19K21875. A. N. is supported by JSPS KAKENHI Grant Nos. JP19H01894 and JP20H04726 and by Research Grants from Inamori Foundation.

\end{document}